\documentclass[prb,aps,twocolumn,showpacs]{revtex4-2} 
\usepackage{amsmath,amssymb,amsthm}
\usepackage{graphicx}
\usepackage{subfigure}
\usepackage{amsmath}
\usepackage{amsfonts,amssymb}
\usepackage{bm}
\usepackage{float}
\usepackage{color}
\usepackage{sidecap}
\allowdisplaybreaks
\usepackage{soul}
\setstcolor{red}
\usepackage[normalem]{ulem}
\usepackage{hyperref}
\hypersetup{colorlinks=true,breaklinks,urlcolor=blue,linkcolor=blue,citecolor=blue}

\begin{document}

\title{Topological Anderson insulators and reentrant topological transitions in a quasiperiodic long-range Su-Schrieffer-Heeger model}

\author{Fang-Ming Meng}
\author{Qi-Bo Zeng}
\email{zengqibo@cnu.edu.cn}
\affiliation{Department of Physics, Capital Normal University, Beijing 100048, China}

\begin{abstract}
We study a one-dimensional long-range Su-Schrieffer-Heeger model with third-nearest-neighbor hopping and subject to quasiperiodic disorder. In the clean limit, the model hosts phases characterized by winding numbers $W=-1,0,1$ and $2$. The introduction of quasiperiodic disorder profoundly modifies the phase diagram and induces a series of topological phase transitions. Owing to the competition between topological dimerization and localization, topological Anderson insulating (TAI) phases with different winding numbers emerge and can persist even when the spectral gap becomes nearly closed in the strong-disorder regime. In addition, we uncover multiple reentrant topological phase transitions induced by varying either the quasiperiodic disorder strength or the hopping amplitudes. Remarkably, the system exhibits staircase-like topological Anderson transitions, where the real-space winding number evolves through successive quantized steps with increasing disorder strength. Our results demonstrate that the interplay between long-range hopping and quasiperiodic disorder generates a rich landscape of disorder-induced topological phases and reentrant topological transition phenomena.
\end{abstract}
\maketitle
\date{today}

\section{Introduction}
Topological phases of matter have become one of the most active research areas in condensed-matter physics and beyond, including cold-atom and photonic systems~\cite{Hasan2010RMP,Qi2011RMP,Bansil2016RMP,Cooper2019RMP,Ozawa2019RMP}. Unlike conventional phases characterized by spontaneous symmetry breaking within the Ginzburg-Landau paradigm~\cite{Ginzburg1950,Wen2017RMP}, topological phases are distinguished by global topological properties of quantum states. Over the past decades, a wide variety of topological phases have been discovered, including topological insulators~\cite{Hasan2010RMP,Qi2011RMP,Bansil2016RMP,Bernevig2006Science}, topological superconductors~\cite{Alicea2012PRP,Beenakker2013ARCMP,Elliott2015RMP}, Weyl and Dirac semimetals~\cite{Hasan2017ARCMP,Yan2017ARCMP,Armitage2018RMP}, nodal-line semimetals~\cite{Fang2016CPB,Yang2022AdvPhys}, higher-order topological insulators~\cite{Sitte2012PRL,Zhang2013PRL,Benalcazar2017Science,Schindler2018SciAdv,Yang2024JPCM}, and, more recently, non-Hermitian topological phases~\cite{Gong2018PRX,Bergholtz2021RMP,Banerjee2023JPCM,Xue2026PRL}. A hallmark of topologically nontrivial systems is the emergence of boundary-localized states protected by quantized topological invariants through the bulk-boundary correspondence. These states are generally robust against external perturbations and weak disorder. By tuning system parameters, topological phase transitions can occur between phases characterized by distinct topological invariants.

Disorder plays a central role in determining the properties of topological systems. While sufficiently strong disorder generally drives Anderson localization~\cite{Anderson1958} and destroys topological order, it can also induce nontrivial topology, leading to the emergence of topological Anderson insulators (TAIs)~\cite{Li2009PRL,Groth2009PRL,Xing2011PRB,Jiang2009PRB,Zhang2012PRB,Song2012PRB,Atland2014PRL,Shem2014PRL,Zhang2019PRB,Hsu2020PRB,Velury2021PRB,Zhang2021PRB,Lu2022AnndePhys,Ji2025arxiv,Mannai2026arxiv}. Disorder-induced topological phases have been extensively investigated in the one-dimensional Su-Schrieffer-Heeger (SSH) model with quasiperiodic disorder~\cite{Liu2018PLA,Longhi2020OL,Tang2022PRA,Li2024PRR,Sircar2025PLA,Wang2026arxiv}, and have been experimentally realized in a variety of platforms, including ultracold atoms~\cite{Meier2018Science}, photonic lattices~\cite{Stutzer2018Nature,Liu2020PRL,Cui2022PRL,Nejad2022AM,Chen2024PRL,Ren2024PRL}, acoustic systems~\cite{Gu2023SCPMA}, and electric circuits~\cite{Zhang2021PRL}. More recently, the concept of TAIs has been extended to non-Hermitian systems~\cite{Zhang2020SciChina,Liu2020CPB,Tang2020PRA,Tang2022PRA,Lin2022NatCom,Zeng2026PRB}. Another intriguing consequence of the interplay between disorder and topology is the emergence of reentrant topological phase transitions, in which a system undergoes successive topological and trivial phases upon varying a control parameter~\cite{Tezuka2012PRB,Beugeling2012PRB,Rieder2013PRB,Tezuka2013PRB,Hsu2017PRB,Sugimoto2017PRB,Yang2020PRB,Padhan2024PRB,Kesharpu2025PRB,Cinnirella2025PRB,Li2025PRB,Lu20252025FoP,Wang2025PRA,Sinha2025PRA}. In particular, long-range hopping provides an additional mechanism for engineering topological phases with larger winding numbers. Previous work has shown that disorder can induce TAI phases in long-range SSH models~\cite{Hsu2020PRB}. However, the effects of quasiperiodic disorder acting on different hopping terms, as well as the possibility of disorder-induced reentrant topological transitions in such systems, remain largely unexplored.

In this work, we investigate a one-dimensional long-range SSH model with third-nearest-neighbor hopping and subject to quasiperiodic disorder. We consider disorder acting selectively on the intracell, intercell, or third-nearest-neighbor hopping amplitudes. In the clean limit, the model hosts distinct phases characterized by winding numbers $W=-1,0,1$ and $2$. Upon introducing quasiperiodic disorder, we find that topological Anderson insulating phases with different winding numbers will emerge, whereby zero-energy edge states are induced from regions that are trivial in the absence of disorder. Disorder also drives a sequence of transitions among these phases. The corresponding phase boundaries are determined from the real-space winding numbers, Lyapunov exponents of the zero-energy modes and corroborated by gap-closing points in the energy spectrum. Remarkably, topologically nontrivial phases survive even in the strong-disorder regime where the spectral gap becomes vanishingly small. In addition, we uncover multiple reentrant topological phase transitions driven by either disorder strength or hopping amplitudes. We further identify staircase-like topological Anderson transitions, in which the winding number evolves through successive quantized changes as the disorder strength increases. These results demonstrate the rich interplay between long-range hopping and quasiperiodic disorder, giving rise to unconventional disorder-induced topological phases and reentrant topological transitions.

The remainder of this paper is organized as follows. In Sec.~\ref{Sec2}, we introduce the long-range SSH model with quasiperiodic disorder and briefly describe the Lyapunov-exponent and real-space winding-number formalisms used throughout this work. In Sec.~\ref{Sec3}, we analyze the topological phases and phase diagram of the clean system. In Sec.~\ref{Sec4}, we investigate the emergence of topological Anderson insulating phases and the associated topological phase transitions induced by quasiperiodic disorder in different hopping channels. Finally, Sec.~\ref{Sec5} summarizes our main results.

\section{Model and topological characterization}\label{Sec2}
We consider a one-dimensional long-range Su-Schrieffer-Heeger (SSH) model with third-nearest-neighbor hopping that preserves chiral symmetry. The Hamiltonian is given by
\begin{equation}\label{H}
  \begin{split}
	H = \sum_{j} & \left( t_1 c_{j,A}^\dagger c_{j,B} + t_2 c_{j,B}^\dagger c_{j+1,A} + t_3 c_{j,A}^\dagger c_{j+1,B} \right. \\
	& \left. + t_4 c_{j,B}^\dagger c_{j+2,A} + h.c. \right),
  \end{split}
\end{equation}
where $c_{j,A(B)}^\dagger$ creates a spinless fermion on sublattice $A(B)$ of the $j$th unit cell. The hopping amplitudes $t_1$ and $t_2$ correspond to the nearest-neighbor intracell and intercell couplings, respectively, while $t_3$ and $t_4$ denote the third-nearest-neighbor hopping amplitudes between the two sublattices. Here $h.c.$ represents the Hermitian conjugate. The lattice contains $N$ unit cells, corresponding to a total system size $L=2N$. Throughout this work, all hopping amplitudes are taken to be real. In the limit $t_3=t_4=0$, Eq.~(\ref{H}) reduces to the conventional SSH model, which is topologically nontrivial for $|t_1|<|t_2|$. The inclusion of long-range hopping enriches the topological structure of the system, allowing phases characterized by winding numbers ranging from $-1$ to $2$, as discussed below.

To investigate disorder-induced topological phenomena, we introduce quasiperiodic disorder into one of the hopping channels and replace the corresponding hopping amplitude in Eq.~(\ref{H}) by
\begin{equation}\label{trj}	
	t_{r,j}^\prime = t_r + \lambda \cos (2\pi \alpha j + \phi), \quad r \in \left\{1,2,3,4\right\},
\end{equation}
where $\lambda$ denotes the disorder strength and $\alpha$ is an irrational number. Throughout this work, we set $\alpha=(\sqrt{5}-1)/2$ and $\phi=0$. With the quasiperiodic disorder applied, the system can exhibit a rich variety of topological phases and phase transitions, as will be shown later in this paper.

In the clean system, topological phase transitions are normally accompanied by bulk-gap closings, which is defined as $\Delta E = E_{N+1} - E_{N}$ under periodic boundary conditions. The bulk gap therefore provides a convenient indicator of phase boundaries. In the presence of quasiperiodic disorder, we complement this analysis with the Lyapunov exponent and real-space winding number introduced below.

The topological properties of the system can be characterized through the behavior of the zero-energy edge states. In the topologically nontrivial phase, these states are exponentially localized at the system boundaries, whereas they are absent in the trivial phase. Consequently, topological phase transitions can be identified by analyzing the localization properties of the zero-energy solutions. To this end, we employ the transfer-matrix formalism and calculate the corresponding Lyapunov exponent. Here we take the case where quasiperiodic disorder is applied to the intracell hopping amplitude as an example, i.e., we set $r=1$ in Eq.~(\ref{trj}). The analysis for disorder in other hopping terms proceeds in an similar manner. Substituting the eigenstate $| \psi \rangle = \sum_j a_j | j,A \rangle + b_j | j, B \rangle$ into the stationary Schrödinger equation $H | \psi \rangle = E | \psi \rangle$, we obtain
\begin{equation}\label{SchoEq}
	\left\{
	\begin{array}{ll}
		E a_j &=  t_{1,j} b_j + t_2 b_{j-1} + t_3 b_{j+1} + t_4 b_{j-2}, \\
		E b_j &=  t_{1,j} a_j + t_2 a_{j+1} + t_3 a_{j-1} + t_4 a_{j+2}.
	\end{array}
	\right.
\end{equation}
Setting $E=0$, the coupled equations decouple into independent recursion relations for the $A$ and $B$ sublattices. Introducing the six-component vector
\begin{equation}
	X_j = \begin{pmatrix}
		a_{j+1} \\
		a_j \\
		a_{j-1} \\
		b_j \\
		b_{j-1} \\
		b_{j-2}
	\end{pmatrix},
\end{equation}
with the boundary conditions $a_0=b_0=b_{-1}=0$, the recursion relations can be cast into the transfer-matrix form
\begin{equation}
	X_{j+1} = T_j X_j,
\end{equation}
where the transfer matrix is given by
\begin{equation}\label{TransMat}
	T_j= \begin{pmatrix}
		-\frac{t_2}{t_4} & -\frac{t_{1,j}}{t_4} & -\frac{t_3}{t_4} & 0 & 0 & 0 \\
		1 & 0 & 0 & 0 & 0 & 0 \\
		0 & 1 & 0 & 0 & 0 & 0 \\
		0 & 0 & 0 & -\frac{t_{1,j}}{t_3} & -\frac{t_2}{t_3} & -\frac{t_4}{t_3} \\
		0 & 0 & 0 & 1 & 0 & 0 \\
		0 & 0 & 0 & 0 & 1 & 0
	\end{pmatrix}.
\end{equation}

The localization properties of the zero-energy modes can then be characterized by the Lyapunov exponent. Following Ref.~\cite{Song2014PRB}, we define
\begin{equation}
	\Gamma = \lim_{N \rightarrow \infty } \left[ T_1^\dagger T_2^\dagger \cdots T_N^\dagger T_N \cdots T_2 T_1 \right]^{1/2N}.
\end{equation}
The matrix $\Gamma$ possesses six eigenvalues, which can be written as $\left\{ e^{\gamma_i}:i=1-6 \right\}$. The localization length is determined as $\xi=\gamma_{LE}^{-1}$, with $\gamma_{LE}$ being the smallest positive Lyapunov exponent among the $\gamma_i$. A finite Lyapunov exponent $(\gamma_{LE}>0)$ indicates exponentially localized zero-energy modes with a finite localization length, whereas $\gamma_{LE}=0$ corresponds to delocalized or critical zero-energy states. Consequently, the vanishing of $\gamma_{LE}$ signals a divergence of the localization length and provides an effective criterion for identifying topological phase transitions.

To further characterize the topological phases in the presence of quasiperiodic disorder, we compute the real-space winding number under open boundary conditions~\cite{Shem2014PRL}. The Hamiltonian preserves chiral symmetry, $SHS^{-1}=-H$, with $S=\sigma_z \otimes \mathcal{I}$, where $\sigma_z$ is the Pauli matrix and $\mathcal{I}$ is the $N$-dimensional identity matrix. Let $| \psi^{\pm}_{n} \rangle$ denotes the eigenstates of Hamiltonian in Eq.~(\ref{H}), which  satisfies the Schr\"odinger equation $H | \psi^{\pm}_{n} \rangle = \pm E_n | \psi^{\pm}_{n} \rangle$. Here the superscripts $\pm$ label the positive- and negative-energy sectors of the spectrum. The open-boundary $Q$ matrix is defined as $Q = \sum_n \left( | \psi^{+}_{n} \rangle \langle \psi^{+}_{n} | - | \psi^{-}_{n} \rangle \langle \psi^{-}_{n} | \right)$. The real-space winding number is then given by
\begin{equation}\label{W}
	W = \frac{1}{2L^\prime} \mathrm{Tr^\prime} \left( SQ [Q, X] \right),
\end{equation} 
where $X$ is the position operator and $\mathrm{Tr}^\prime$ denotes the trace over a central region of length $L^\prime$. Throughout this work, we choose $L^\prime=N$. The winding number $W$ serves as a topological invariant that distinguishes the different topological phases of the system. Unlike the conventional momentum-space winding number, the real-space formulation remains well defined in the absence of translational symmetry and therefore provides a unified characterization of both clean and disordered systems.

\section{Topological Phases in the Clean Limit}\label{Sec3}
\begin{figure}[t]
	\includegraphics[width=3.0in]{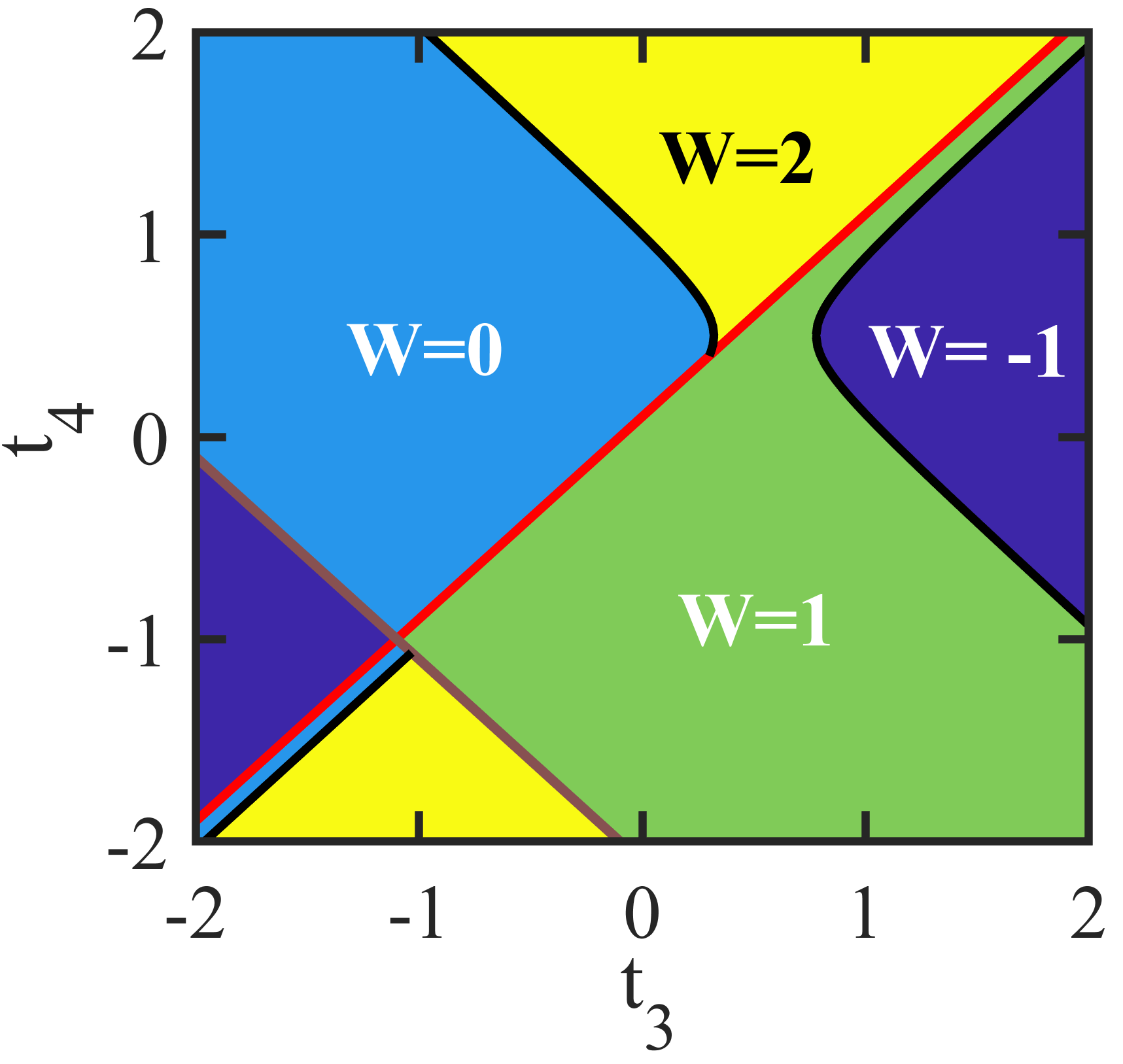}
	\caption{(Color online) Phase diagram of the clean long-range SSH model in the $t_3$-$t_4$ plane for $t_1=1$ and $t_2=1.1$. The colored regions correspond to phases with winding numbers $W=-1$, $0$, $1$, and $2$. The solid lines denote the phase boundaries determined from the bulk-gap closing conditions.}
	\label{fig1}
\end{figure}

We first examine the topological phases of the long-range SSH model in the absence of quasiperiodic disorder. For $t_3=t_4=0$, Eq.~(\ref{H}) reduces to the conventional SSH model, which hosts a topological phase with one zero-energy edge state at each boundary when $|t_1|<|t_2|$. This phase is characterized by a winding number $W=1$. The inclusion of long-range hopping substantially enriches the topological structure of the system and gives rise to phases with different winding numbers. Figure~\ref{fig1} presents the phase diagram in the $t_3$-$t_4$ plane for $t_1=1$ and $t_2=1.1$. The colored regions correspond to phases with winding numbers $W=-1,0,1,$ and $2$, obtained from the real-space winding number defined in Sec.~\ref{Sec2}. The existence of the $W=2$ phase demonstrates that long-range hopping can generate topological phases with higher winding numbers than those accessible in the conventional SSH model. By tuning the $t_3$ and $t_4$, the system undergoes a sequence of topological phase transitions between phases characterized by different winding numbers.

Since translational symmetry is preserved in the clean system, the Hamiltonian can be transformed into momentum space via a Fourier transformation, yielding
\begin{equation}
	\begin{aligned}
	& H(k) =  \\
	& \begin{pmatrix}
		0 & t_1 + t_2 e^{-ik} + t_3 e^{ik} + t_4 e^{-2ik} \\
		t_1 + t_2 e^{ik} + t_3 e^{-ik} + t_4 e^{2ik} & 0
	\end{pmatrix},
	\end{aligned}
\end{equation}
where $k \in [-\pi, \pi]$. The energy spectrum consists of two bands, and topological phase transitions occur when the bulk gap closes. The gap-closing conditions are obtained from $\det[H(k)]=0$, leading to
\begin{equation}
	\begin{aligned}
		t_1 - t_2 - t_3 + t_4 &= 0, \\
		t_1 + t_2 + t_3 + t_4 &= 0, \\
		t_4^2 - t_3^2 - t_1 t_4 + t_2 t_3 &= 0 \quad \text{with} \quad \left| \frac{t_3-t_2}{2 t_4} \right| \leq 1.
    \end{aligned}
\end{equation}
Substituting $t_1=1$ and $t_2=1.1$ yields the phase boundaries shown by the solid lines in Fig.~\ref{fig1}. The analytical phase boundaries are in excellent agreement with the phase diagram obtained from the real-space winding number.

\begin{figure}[t]
	\includegraphics[width=3.2in]{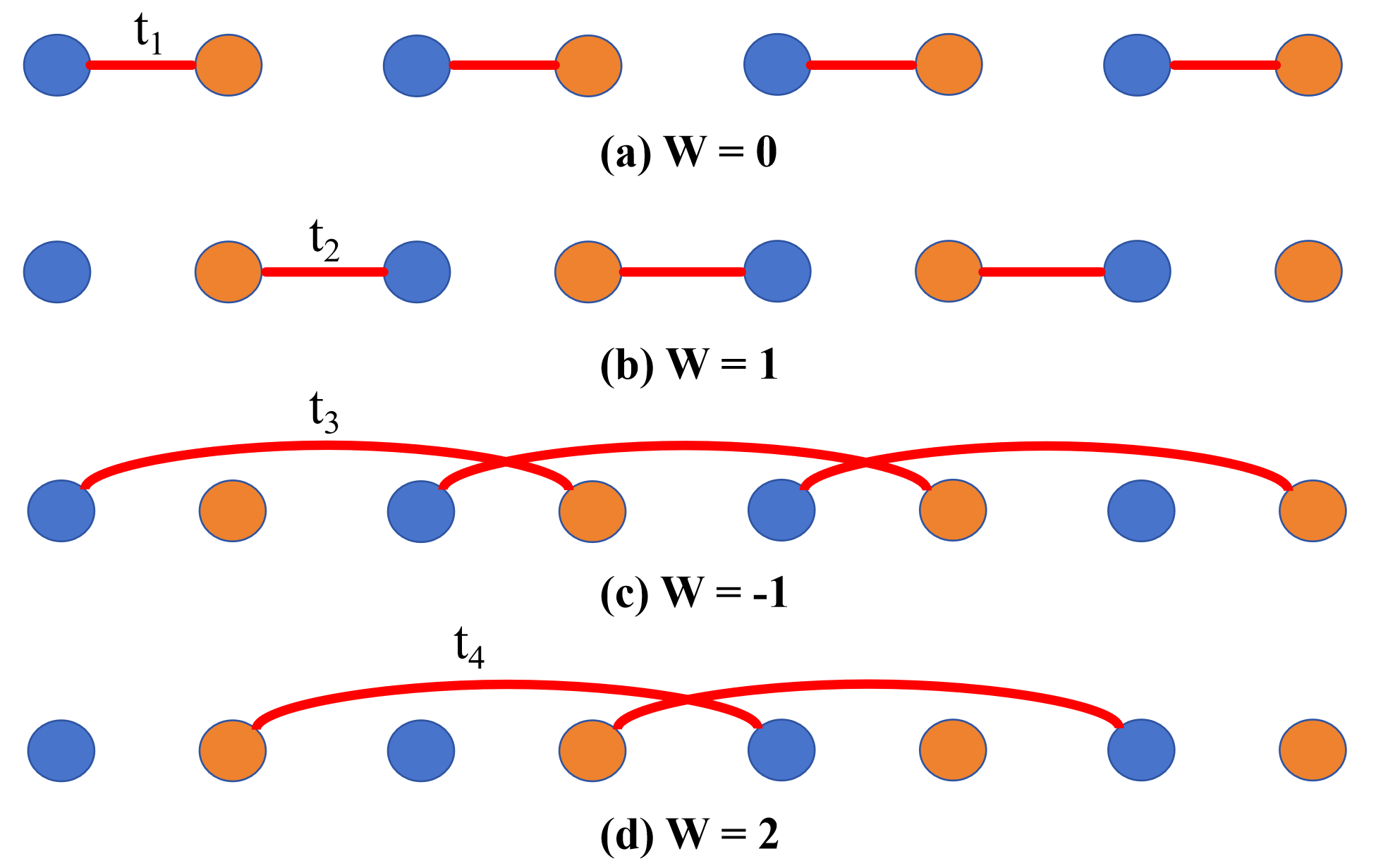}
	\caption{(Color online) Schematic illustration of the dimerization configurations and the associated zero-energy edge states in the long-range SSH model. The blue and orange circles represent the two sublattices of each unit cell, while the red lines denote the dominant hopping amplitudes. Panels (a)–(d) correspond to the phases with winding numbers $W=0$, $1$, $-1$, and $2$, respectively. The number and sublattice character of the unpaired boundary sites determine the topological invariant and the corresponding zero-energy edge modes.}
	\label{fig2}
\end{figure}

The origin of these topological phases can be understood from the dimerization patterns illustrated in Fig.~\ref{fig2}. Each unit cell contains two sublattice sites, represented by the blue and orange circles. Different hopping amplitudes favor different pairing configurations between the sublattice sites, leading to distinct edge-state structures and topological invariants. When the intracell hopping $t_1$ dominates, the two sites within each unit cell form dimers, leaving no unpaired sites at the boundaries [Fig.~\ref{fig2}(a)]. The system is therefore topologically trivial with $W=0$. In contrast, when the intercell hopping $t_2$ is dominant, neighboring unit cells are dimerized, leaving one unpaired site at each end of the chain [Fig.~\ref{fig2}(b)]. This configuration supports one zero-energy edge state at each boundary and corresponds to the topological phase with $W=1$.

A similar mechanism applies when the long-range hopping terms become dominant. For sufficiently large $t_3$, the lattice is dimerized as shown in Fig.~\ref{fig2}(c), again leaving one unpaired site at each boundary. However, the unpaired sites now belong to the opposite sublattice compared with the $W=1$ phase, giving rise to a winding number $W=-1$. When the third-nearest-neighbor hopping $t_4$ dominates, the dimerization pattern shown in Fig.~\ref{fig2}(d) leaves two unpaired sites at each boundary. Consequently, the system supports two zero-energy edge states per edge and realizes a topological phase with winding number $W=2$.

The above dimerization picture provides an intuitive understanding of the clean-limit phase diagram. It also suggests that introducing quasiperiodic disorder into different hopping channels may substantially modify the competition among these pairing configurations, thereby inducing a variety of disorder-driven topological phase transitions. This possibility is explored in the following section.

\section{Topological Anderson Insulators and Reentrant Topological Transitions}\label{Sec4}

We next explore the effects of quasiperiodic disorder on the long-range SSH model. To elucidate the role of different hopping processes, quasiperiodic disorder is introduced separately into each of the four hopping channels. This enables a systematic investigation of how disorder modifies the clean-limit topological phases and induces new disorder-driven phenomena. By tuning the disorder strength or hopping amplitudes, the system will exhibit topological Anderson insulating phases, multiple reentrant topological transitions, and staircase-like changes of the winding number.

\subsection{System with disorder in $t_1$}
We first consider quasiperiodic disorder in the intracell hopping amplitude $t_1$. Previous studies have shown that, in the conventional SSH model, disorder in the intracell hopping can induce topologically nontrivial phases from an initially trivial regime, giving rise to topological Anderson insulators (TAIs)~\cite{Tang2022PRA,Wang2025PRA}. A similar mechanism operates in the long-range SSH model considered here, although the resulting phase diagram is substantially richer owing to the presence of long-range hopping.

\begin{figure}[t]
	\includegraphics[width=3.4in]{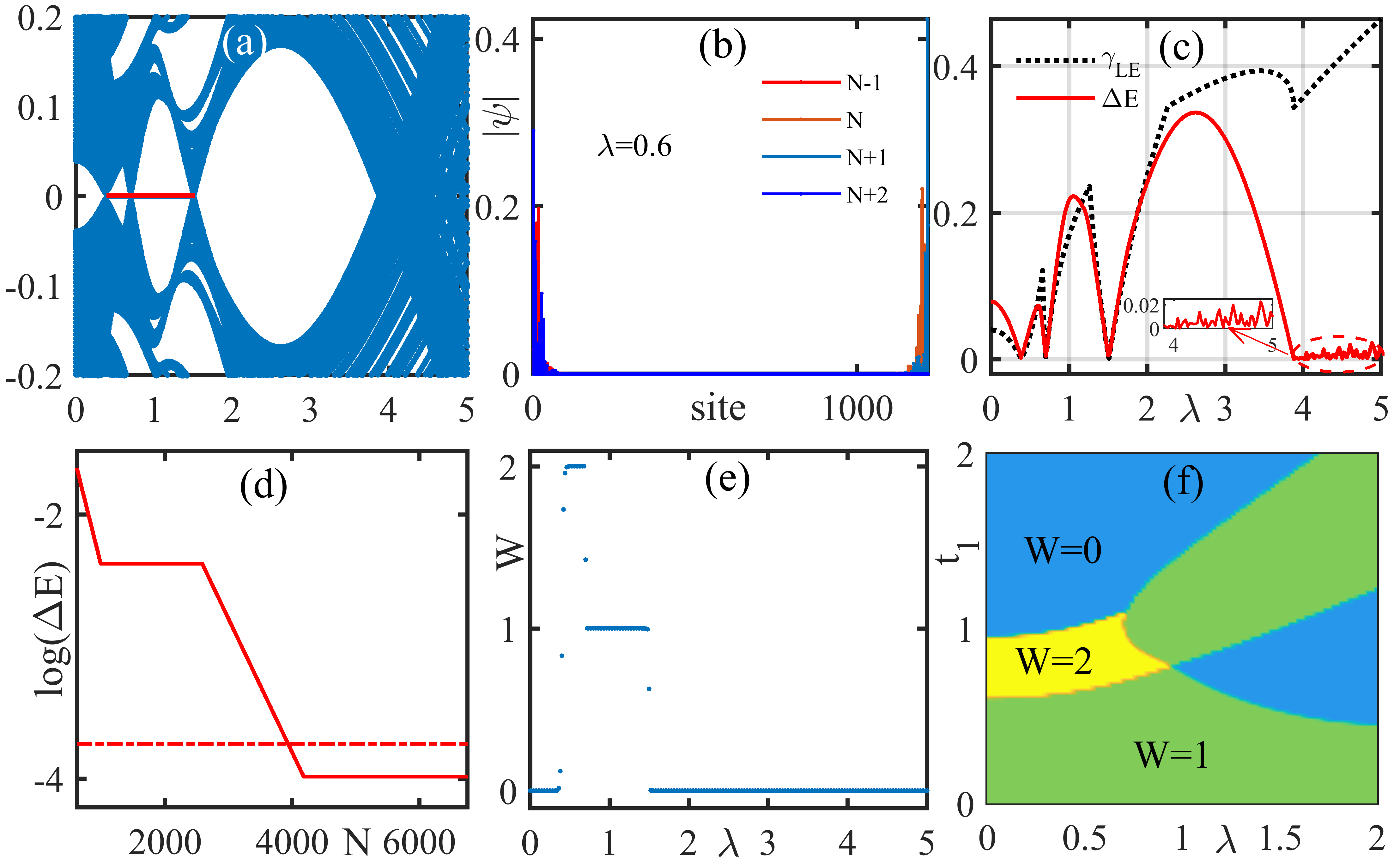}
	\caption{(Color online) Topological phases of the long-range SSH model with quasiperiodic disorder in the intracell hopping $t_1$. (a) Energy spectrum as a function of the disorder strength $\lambda$ under open boundary conditions. The red line represents the zero-energy edge modes. (b) Spatial distributions of the four zero-energy edge modes at $\lambda=0.6$. (c) Lyapunov exponent of the zero-energy modes $\gamma_{LE}$ (black dotted line) and the bulk gap $\Delta E$ around zero energy (red solid line) as functions of $\lambda$. The inset enlarges the region enclosed by the red dashed ellipse. (d) Logarithm of the bulk gap, $\log(\Delta E)$, as a function of system size for $\lambda=4.06$ (red dot-dashed line) and $\lambda=4.88$ (red solid line). (e) Real-space winding number $W$ as a function of $\lambda$. (f) Topological phase diagram in the $t_1$--$\lambda$ plane. The numbers shown in the colored regions denote the corresponding winding numbers. System parameters: $t_1=1$, $t_2=1.1$, $t_3=0.2$, $t_4=0.7$ and $L=2N=1220$.}
	\label{fig3}
\end{figure}

Figure~\ref{fig3}(a) shows the energy spectrum under open boundary conditions as a function of the disorder strength $\lambda$ for $t_1=1$, $t_2=1.1$, $t_3=0.2$, and $t_4=0.7$. At $\lambda=0$, the spectrum is gapped near $E=0$. The gap is closed and reopened at $\lambda \approx 0.38$, accompanied by the appearance of four zero-energy edge modes, as indicated by the red line. The spatial distributions of the zero modes are shown in Fig.~\ref{fig3}(b). Upon further increasing $\lambda$, the bulk gap closes and reopens near $\lambda \approx 0.7$, and there are only two zero-energy modes left in the gap. A third gap-closing transition occurs at $\lambda \approx 1.51$, beyond which the zero modes disappear. The corresponding edge states are localized at the two ends of the chain and signal the existence of topologically nontrivial phases. In the regime $0.38 \lesssim \lambda \lesssim 0.7$ hosts four zero modes, whereas two zero modes are present for $0.7 \lesssim \lambda \lesssim 1.51$.

The bulk-gap evolution is summarized in Fig.~\ref{fig3}(c). The gap-closing points in the bulk gap,  which is indicated by the red solid curve here, coincide with the topological phase transitions. For $\lambda \gtrsim 4$, the bulk gap becomes extremely small but remains finite. To verify this behavior, Fig.~\ref{fig3}(d) presents $\log(\Delta E)$ as a function of system size for representative values of $\lambda$. Rather than vanishing continuously with increasing system size, the gap saturates at a finite value, indicating that the strongly disordered phase remains weakly gapped. This behavior originates from the quasiperiodic hopping term $t_{1,j}^{\prime}$, whose amplitude does not vanish identically for irrational modulation. In the strong-disorder limit, $t_{1,j}^{\prime}$ dominates in determining the energy spectrum, leading to the fluctuating and small bulk gaps. Similar nearly gapless yet finite-gap regimes also appear when quasiperiodic disorder is introduced into the other hopping terms.

To determine the phase boundaries, we compute the Lyapunov exponent $\gamma_{LE}$ of the zero-energy states. As shown in Fig.~\ref{fig3}(c), the zeros of $\gamma_{LE}$ coincide precisely with the bulk-gap closing points, confirming the locations of the topological transitions. The corresponding real-space winding number is plotted in Fig.~\ref{fig3}(e). Starting from the trivial phase with (W=0), the winding number jumps to $W=2$ at $\lambda \approx 0.38$, indicating the emergence of two zero-energy edge modes at each boundary. With further increasing disorder strength, the winding number decreases sequentially from $W=2$ to $W=1$ and eventually to $W=0$. The transition points obtained from the winding number agree quantitatively with those determined from the bulk-gap and Lyapunov-exponent analyses.

The resulting phase diagram in the $t_1$-$\lambda$ plane is shown in Fig.~\ref{fig3}(f) for $t_2=1.1$, $t_3=0.2$ and $t_4=0.7$. Three topological phases characterized by $W=0$, $1$, and $2$ are observed. Quasiperiodic disorder induces TAI phases with both $W=1$ and $W=2$, demonstrating that long-range hopping significantly enriches the disorder-induced topological phases compared with the conventional SSH model. Moreover, with appropriate parameters, the winding number changes sequentially according to $2 \rightarrow 1 \rightarrow 0$, revealing a staircase-like topological Anderson transition driven by quasiperiodic disorder.

\begin{figure}[t]
	\includegraphics[width=3.2in]{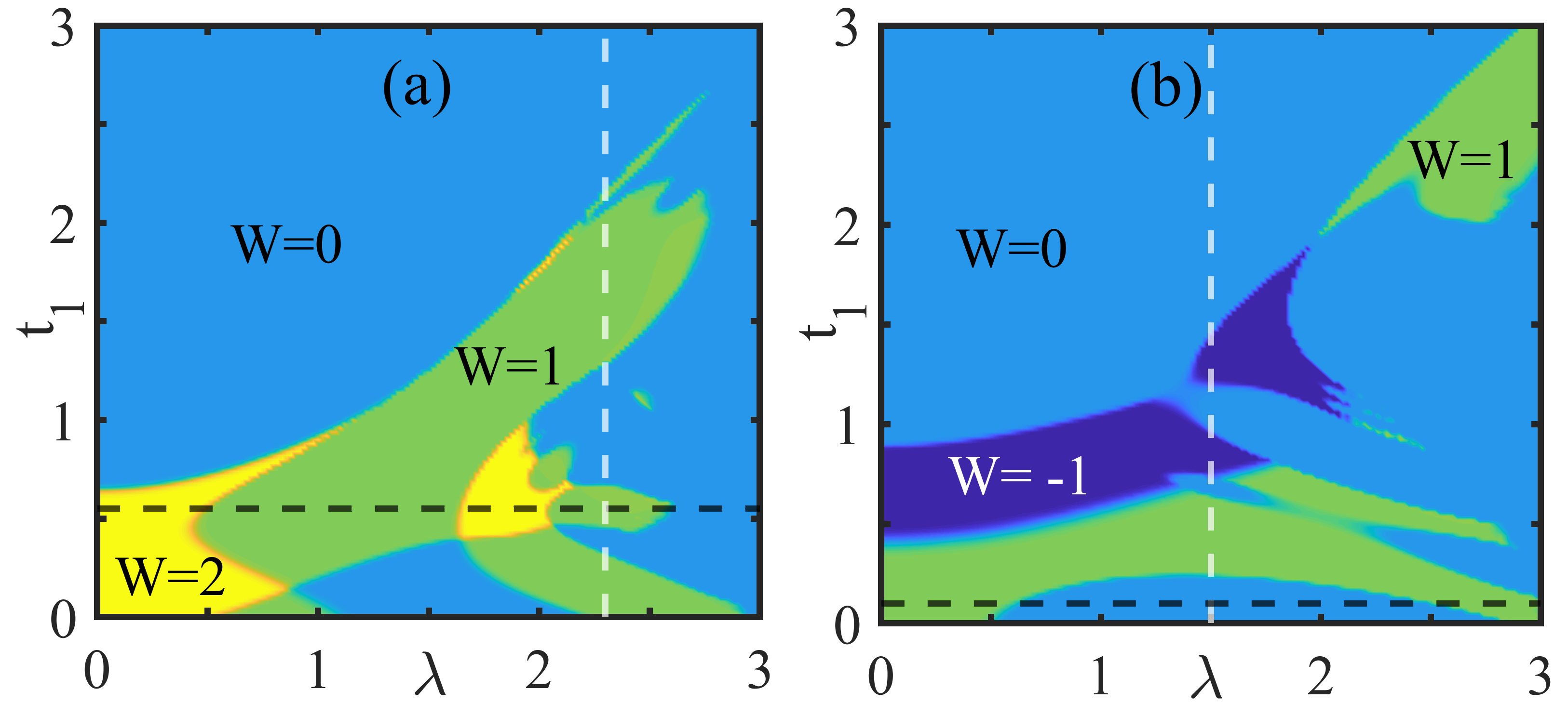}
	\caption{(Color online) Topological phase diagrams in the $t_1$-$\lambda$ plane for the long-range SSH model with quasiperiodic disorder in the intracell hopping $t_1$. Different colors denote phases with different winding numbers. (a) Phase diagram for $t_2=0.2$, $t_3=0.8$, and $t_4=1.1$. The black dashed line marks the cut at $t_1=0.55$, while the white dashed line corresponds to $\lambda=2.3$, which exhibits multiple reentrant transitions in the system. (b) Phase diagram for $t_2=0.7$, $t_3=1.1$, and $t_4=0.9$. The black and white dashed lines indicate the cuts at $t_1=0.1$ and $\lambda=1.5$, respectively, along which additional reentrant topological transitions occur.}
	\label{fig4}
\end{figure}

Figure~\ref{fig4} presents two additional phase diagrams in the $t_1$-$\lambda$ plane for different parameter sets, revealing even richer disorder-induced topological phenomena. Along the black dashed line at $t_1=0.55$ in Fig.~\ref{fig4}(a), the winding number evolves according to $W=2 \rightarrow 1 \rightarrow 2 \rightarrow 1 \rightarrow 0 $ as the disorder strength increases. This sequence demonstrates reentrant topological transitions between the $W=2$ and $W=1$ phases. Reentrant behavior can also be induced by varying $t_1$. Along the white dashed line in Fig.~\ref{fig4}(a), the system undergoes repeated transitions between the $W=1$ and $W=0$ phases, providing another example of multiple reentrant topological transitions. Similar phenomena are observed in Fig.~\ref{fig4}(b), where the black and white dashed lines indicate reentrant transitions between the $W=1$ and $W=0$ phases and between the $W=0$ and $W=-1$ phases, respectively.

These results demonstrate that quasiperiodic disorder substantially enriches the topological phase structure of the long-range SSH model. Depending on the hopping parameters and disorder strength, the system can host topological Anderson insulating phases with higher winding numbers, staircase-like topological Anderson transitions, and multiple reentrant topological transitions between distinct topological phases.

\subsection{System with disorder in $t_2$}

\begin{figure}[t]
	\includegraphics[width=3.2in]{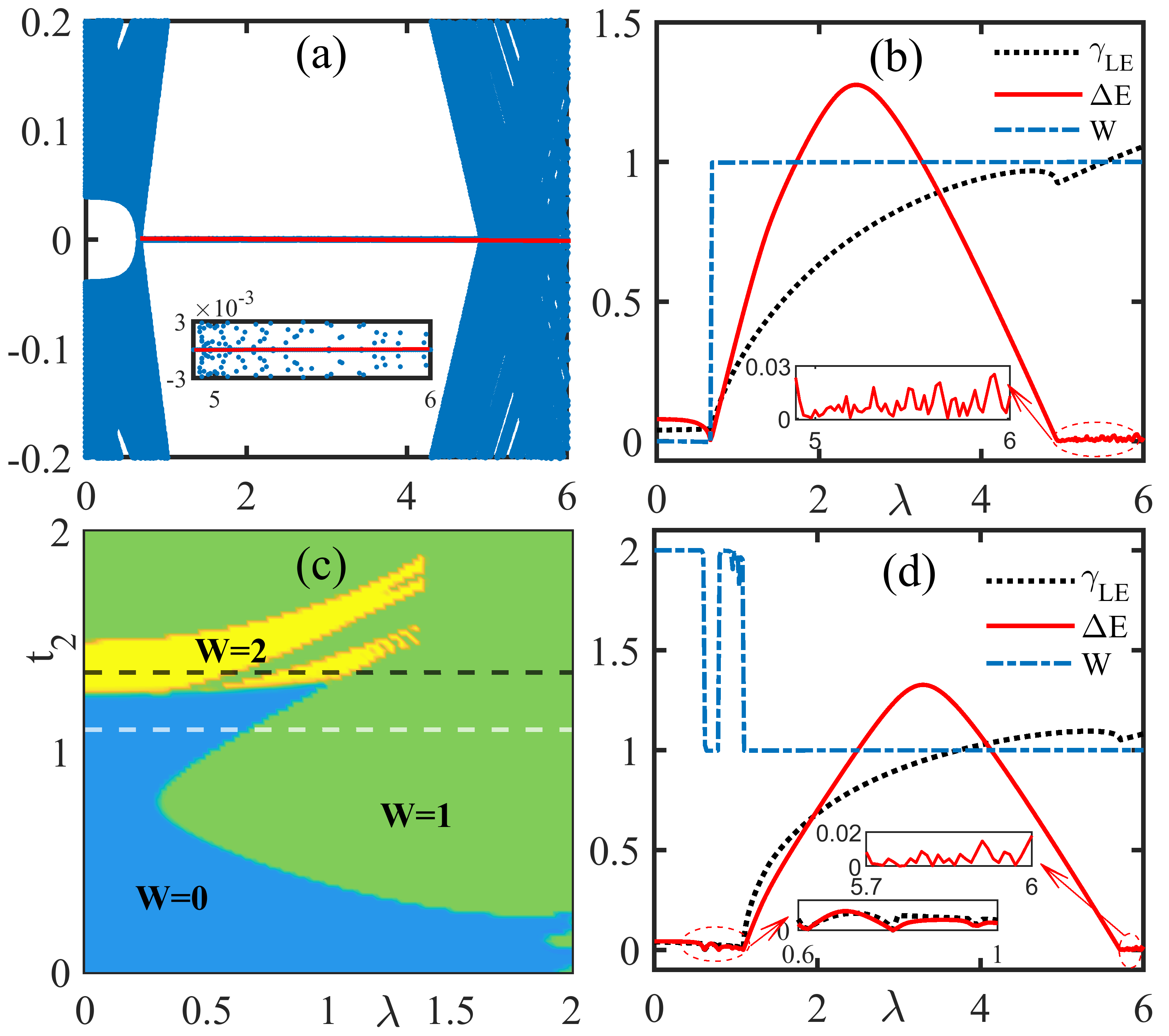}
	\caption{(Color online) Topological phases of the long-range SSH model with quasiperiodic disorder in the intercell hopping $t_2$. (a) Energy spectrum as a function of the disorder strength $\lambda$. The red line denotes the zero-energy edge modes and the inset enlarges the spectrum for $5<\lambda<6$. (b) Lyapunov exponent $\gamma_{LE}$ (black dotted line), bulk gap $\Delta E$ (red solid line), and real-space winding number $W$ (blue dot-dashed line) as functions of $\lambda$. The inset enlarges the region enclosed by the red dashed ellipse. The parameters are $t_1=1$, $t_2=1.1$, $t_3=0.2$, $t_4=0.7$, and $L=2N=1220$. (c) Topological phase diagram in the $t_2$-$\lambda$ plane. The white dashed line corresponds to the parameter cut analyzed in panels (a) and (b). The black dashed line denotes $t_2=1.36$, along which the system undergoes reentrant topological phase transitions. (d) Lyapunov exponent $\gamma_{LE}$, bulk gap $\Delta E$, and winding number $W$ as functions of $\lambda$ for $t_2=1.36$, illustrating the reentrant sequence $W=2 \rightarrow 1 \rightarrow 2 \rightarrow 1$.}
	\label{fig5}
\end{figure}

We next consider quasiperiodic disorder in the intercell hopping amplitude $t_2$, corresponding to $r=2$ in Eq.~(\ref{trj}), such that $t_{2,j}^{\prime}=t_2+\lambda\cos(2\pi\alpha j)$. In contrast to the case of disorder in $t_1$, the resulting topological phase diagram exhibits a different structure. From the dimerization picture shown in Fig.~\ref{fig2}(b), one expects that sufficiently strong disorder drives the system toward a phase dominated by the intercell hopping $t_{2,j}^{\prime}$, which is characterized by a winding number $W=1$.

Figures~\ref{fig5}(a) and \ref{fig5}(b) show the energy spectrum and the corresponding topological features for $t_1=1$, $t_2=1.1$, $t_3=0.2$, and $t_4=0.7$. In the weak-disorder regime, there is a small energy gap near zero energy and the system remains topologically trivial. As the disorder strength increases, the gap closes and reopens, where zero-energy edge modes emerges within the gap. This signals the formation of a topological Anderson insulating phase. For larger values of $\lambda$, the bulk gap becomes extremely small and appears to close. However, a closer inspection of the spectrum [see the inset of Fig.~\ref{fig5}(a)] reveals that the zero-energy edge modes persist. This observation is confirmed by the bulk gap $\Delta E$, shown by the red curve in Fig.~\ref{fig5}(b), which remains finite although very small in the interval $5 \lesssim \lambda \lesssim 6$. Similar to the case of disorder in $t_1$, the gap exhibits strong fluctuations but does not vanish identically. Consistently, neither the Lyapunov exponent nor the real-space winding number displays any signature of a topological transition in this regime, as shown by the black dotted and blue dashed curves in Fig.~\ref{fig5}(b).

The complete phase diagram in the $t_2$-$\lambda$ plane is presented in Fig.~\ref{fig5}(c). Besides the disorder-induced TAI phase observed along the cut $t_2=1.1$, the phase diagram also exhibits multiple reentrant topological transitions. A representative example is highlighted by the black dashed line at $t_2=1.36$. In the clean limit, the system resides in the $W=2$ phase. As the disorder strength increases, the winding number first decreases from $W=2$ to $W=1$. Upon further increasing $\lambda$, the system reenters the $W=2$ phase before undergoing a second transition back to the $W=1$ phase, which remains stable in the strong-disorder regime. The corresponding evolution of the winding number is shown in Fig.~\ref{fig5}(d). The transition points coincide with the zeros of the Lyapunov exponent and the bulk-gap closing points, confirming their topological origin.

In the strong-disorder limit, as well as for sufficiently large values of $t_2$, the phase with $W=1$ occupies a substantial portion of the phase diagram. This behavior can be understood from the dimerization pattern illustrated in Fig.~\ref{fig2}(b), where the intercell hopping dominates the lattice structure. Therefore, quasiperiodic disorder in the intercell hopping not only induces topological Anderson insulating phases but also gives rise to multiple reentrant topological transitions between phases with different winding numbers.

\subsection{System with Disorder in $t_3$ or $t_4$}

\begin{figure}[t]
	\includegraphics[width=3.2in]{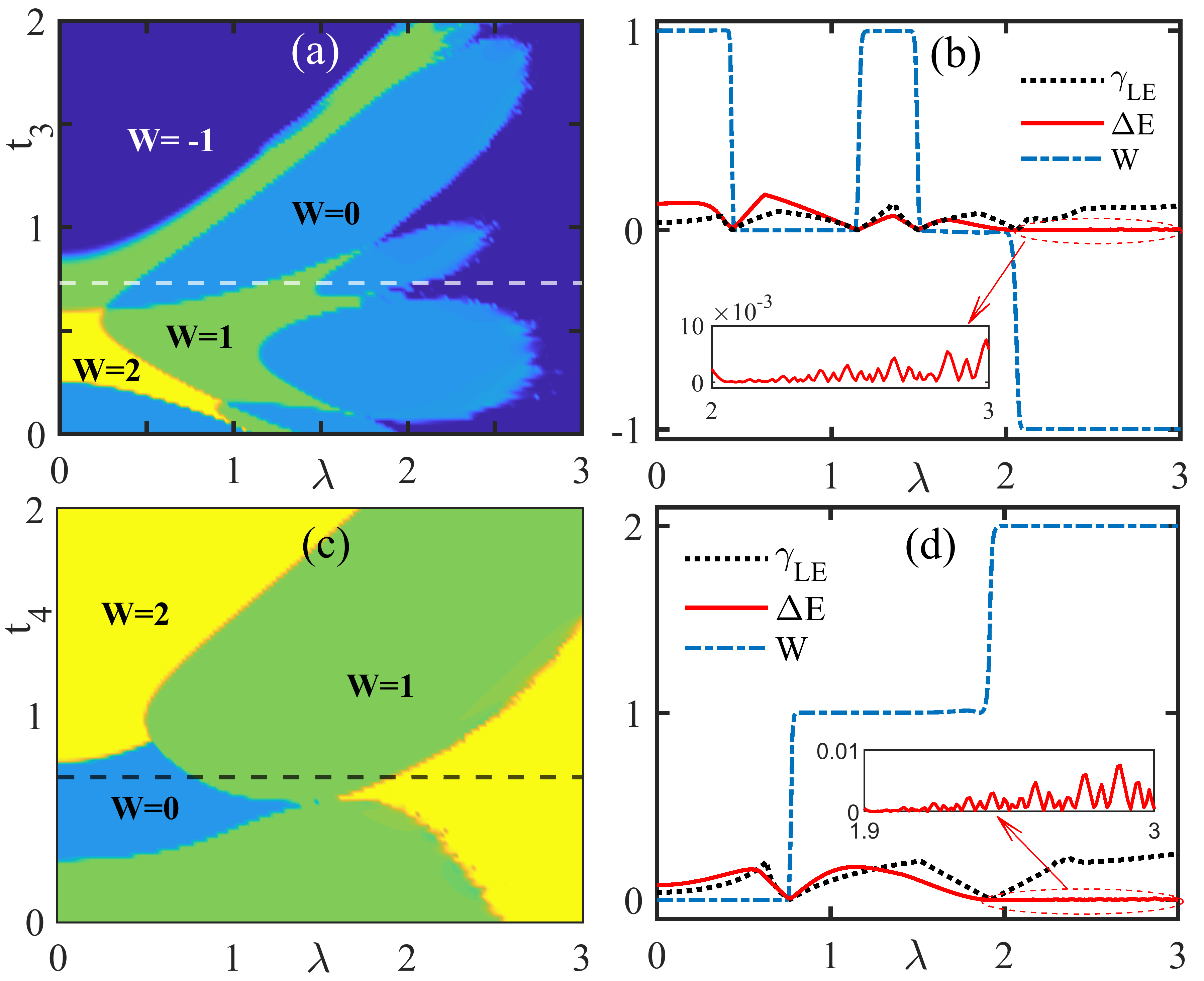}
	\caption{(Color online) Topological phases of the long-range SSH model with quasiperiodic disorder in the long-range hopping channels $t_3$ and $t_4$. (a) Topological phase diagram in the $t_3$--$\lambda$ plane for $t_1=1$, $t_2=1.1$, and $t_4=0.7$. Different colors correspond to phases with winding numbers ranging from $W=-1$ to $W=2$. The white dashed line indicates the cut at $t_3=0.73$, along which multiple reentrant topological transitions occur. (b) Lyapunov exponent $\gamma_{LE}$ (black dotted line), bulk gap $\Delta E$ (red solid line), and real-space winding number $W$ (blue dot-dashed line) as functions of $\lambda$ along the cut shown in (a). (c) Topological phase diagram in the $t_4$--$\lambda$ plane for $t_1=1$, $t_2=1.1$, and $t_3=0.2$. The black dashed line denotes the cut at $t_4=0.7$. (d) Corresponding $\gamma_{LE}$, $\Delta E$, and $W$ as functions of $\lambda$, illustrating a staircase topological Anderson transition with increasing winding number. The insets in (b) and (d) enlarge the regions enclosed by the red ellipses. The system size is $L=2N=1220$.}
	\label{fig6}
\end{figure}

We next consider quasiperiodic disorder in the long-range hopping amplitudes $t_3$ and $t_4$. Figure~\ref{fig6}(a) presents the phase diagram for quasiperiodic disorder in $t_3$. In the regime of large $t_3$ or strong disorder, the system is predominantly governed by the modulated hopping $t_3^{\prime}$ and approaches the dimerization pattern shown in Fig.~\ref{fig2}(c). Consequently, the phase with winding number $W=-1$ occupies a substantial portion of the phase diagram. In addition, phases with winding numbers ranging from $W=-1$ to $W=2$ emerge as $t_3$ and $\lambda$ are varied.

A particularly striking feature is the occurrence of multiple reentrant topological transitions. Along the white dashed line in Fig.~\ref{fig6}(a), the system initially resides in the (W=1) phase. As the disorder strength increases, it undergoes the sequence $W=1 \rightarrow 0 \rightarrow 1 \rightarrow 0 \rightarrow -1$. The corresponding Lyapunov exponent, bulk gap, and real-space winding number are shown in Fig.~\ref{fig6}(b). The transition points determined from these three quantities are in excellent agreement. Similar reentrant behavior can also be induced by tuning the long-range hopping amplitude $t_3$ at fixed disorder strength. For example, at $\lambda=1.5$, the system undergoes multiple transitions between the $W=0$ and $W=1$ phases as $t_3$ increases, whereas at $\lambda=2.2$ repeated transitions between the $W=-1$ and $W=0$ phases are observed.

Another notable feature of Fig.~\ref{fig6}(a) is the staircase evolution of the winding number. For representative parameters such as $t_3=0.5$, increasing the disorder strength drives the system through a sequence of topological phases with progressively decreasing winding number, $W=2 \rightarrow 1 \rightarrow 0 \rightarrow -1$. This behavior constitutes a staircase-like topological Anderson transition, characterized by successive quantized changes of the winding number as the disorder strength increases.

The phase diagram for quasiperiodic disorder in the $t_4$ hopping channel is shown in Fig.~\ref{fig6}(c) for $t_1=1$, $t_2=1.1$, and $t_3=0.2$. In the limit of large $t_4$ or strong disorder, the hopping $t_4^{\prime}$ dominates the lattice structure, leading to the $W=2$ phase associated with the dimerization pattern in Fig.~\ref{fig2}(d). Along the cut $t_4=0.7$, indicated by the black dashed line, the system is initially topologically trivial. As the disorder strength increases, it first enters the $W=1$ phase and subsequently the $W=2$ phase. The corresponding evolution of the winding number is shown in Fig.~\ref{fig6}(d), where the transition points coincide with the bulk-gap closings and the zeros of the Lyapunov exponent. In this case, quasiperiodic disorder drives a staircase transition with increasing winding number, $W=0 \rightarrow 1 \rightarrow 2$.

The results for disorder in $t_3$ and $t_4$ demonstrate the rich interplay between long-range hopping and quasiperiodic disorder. Depending on the hopping channel and parameter regime, the system can exhibit staircase-like topological Anderson transitions, multiple reentrant topological transitions, and disorder-induced phases with winding numbers ranging from $W=-1$ to $W=2$.

\section{Summary}\label{Sec5}

In summary, we have investigated the topological phases of a one-dimensional long-range SSH model with quasiperiodic disorder in the hopping amplitudes. In the clean limit, the model supports topological phases characterized by winding numbers ranging from $W=-1$ to $W=2$. By introducing quasiperiodic disorder into different hopping channels, we uncover a rich variety of disorder-induced topological phenomena.

We show that quasiperiodic disorder can drive transitions from topologically trivial phases to nontrivial phases, giving rise to topological Anderson insulators with different winding numbers. Remarkably, some of these topological phases persist deep in the strong-disorder regime, where the bulk gap becomes extremely small but remains finite. In addition, the interplay between long-range hopping and quasiperiodic disorder leads to multiple reentrant topological transitions as well as staircase-like topological Anderson transitions, in which the winding number changes successively between different quantized values. The topological phase boundaries are consistently identified through the real-space winding number, the Lyapunov exponent of the zero-energy modes, and the bulk-gap closing points.

Our results demonstrate that quasiperiodic disorder and long-range hopping cooperate to generate a rich landscape of topological phases and transitions in one-dimensional systems. These findings provide new insights into disorder-induced topology and may stimulate further studies of topological Anderson phases and reentrant topological phenomena in long-range and quasiperiodic lattices.

\begin{acknowledgments}
	This work is supported by the National Natural Science Foundation of China under the Grant No. 12204326.
\end{acknowledgments}


\begin{thebibliography}{}
\bibitem{Hasan2010RMP}{M. Z. Hasan and C. L. Kane, Colloquium: Topological insulators, \href{https://doi.org/10.1103/RevModPhys.82.3045}{Rev. Mod. Phys. \textbf{82,} 3045 (2010)}.}
	
\bibitem{Qi2011RMP}{X.-L. Qi and S.-C. Zhang, Topological insulators and superconductors, \href{https://doi.org/10.1103/RevModPhys.83.1057}{Rev. Mod. Phys. \textbf{83,} 1057 (2011)}.}
	
\bibitem{Bansil2016RMP}{A. Bansil, H. Lin, and T. Das, Colloquium: Topological band theory, \href{https://doi.org/10.1103/RevModPhys.88.021004}{Rev. Mod. Phys. \textbf{88,} 021004 (2016)}.}
	
\bibitem{Cooper2019RMP}{N. R. Cooper, J. Dalibard, and I. B. Spielman, Topological bands for ultracold atoms, \href{https://doi.org/10.1103/RevModPhys.91.015005}{Rev. Mod. Phys. \textbf{91,} 015005 (2019)}.}
	
\bibitem{Ozawa2019RMP}{T. Ozawa, H. M. Price, A. Amo, N. Goldman, M. Hafezi, L. Lu, M. C. Rechtsman, D. Schuster, J. Simon, O. Zilberberg, and I. Carusotto, Topological photonics, \href{https://doi.org/10.1103/RevModPhys.91.015006}{Rev. Mod. Phys. \textbf{91,} 015006 (2019)}.}
	
\bibitem{Ginzburg1950}{V. L. Ginzburg and L. D. Landau, Zh. Eksp. Teor. Fiz. 20, 1064 (1950).}
	
\bibitem{Wen2017RMP}{X.-G. Wen, Colloquium: Zoo of quantum-topological phases of matter, \href{https://doi.org/10.1103/RevModPhys.89.041004}{Rev. Mod. Phys. \textbf{89,} 041004 (2017)}.}
	
\bibitem{Bernevig2006Science}{B. A. Bernevig, T. L. Hughes, and S. C. Zhang, Quantum spin Hall effect and topological phase transition in HgTe quantum wells \href{https://doi.org/10.1126/science.1133734}{Science \textbf{314,} 1757 (2006)}.}

\bibitem{Alicea2012PRP}{J. Alicea, New directions in the pursuit of Majorana fermions in solid state systems, \href{https://doi.org/10.1088/0034-4885/75/7/076501}{Rep. Prog. Phys. \textbf{75,} 076501 (2012)}.}

\bibitem{Beenakker2013ARCMP}{C. W. J. Beenakker, Search for Majorana Fermions in Superconductors, \href{https://doi.org/10.1146/annurev-conmatphys-030212-184337}{Annu. Rev. Condens. Matter Phys. \textbf{4,} 113 (2013)}.}

\bibitem{Elliott2015RMP}{S. R. Elliott and M. Franz, Colloquium: Majorana fermions in nuclear, particle, and solid-state physics, \href{https://doi.org/10.1103/RevModPhys.87.137}{Rev. Mod. Phys. \textbf{87,} 137 (2015)}.}
	
\bibitem{Hasan2017ARCMP}{M. Z. Hasan, S.-Y. Xu, I. Belopolsi, and S.-M. Huang, Discovery of Weyl fermion semimetals and topological Fermi arc states, \href{https://doi.org/10.1146/annurev-conmatphys-031016-025225}{Annu. Rev. Condens. Matter Phys. \textbf{8,} 289 (2017)}.}
	
\bibitem{Yan2017ARCMP}{ B. Yan and C. Felser, Topological materials: Weyl semimetals, \href{https://doi.org/10.1146/annurev-conmatphys-031016-025458}{Annu. Rev. Condens. Matter Phys. \textbf{8,} 337 (2017)}.}
	
\bibitem{Armitage2018RMP}{N. P. Armitage, E. J. Mele, and A. Vishwanath, Weyl and Dirac semimetals in three-dimensional solids, \href{https://doi.org/10.1103/RevModPhys.90.015001}{Rev. Mod. Phys. \textbf{90,} 015001 (2018)}.}
	
\bibitem{Fang2016CPB}{C. Fang, H. Weng, X. Dai, and Z. Fang, Topological nodal line semimetals, \href{https://doi.org/10.1088/1674-1056/25/11/117106}{Chinese Phys. B \textbf{25,} 117106 (2016)}.}

\bibitem{Yang2022AdvPhys}{M.-X. Yang, W. Luo, and W. Chen, Quantum transport in topological nodal-line semimetals, \href{https://doi.org/10.1080/23746149.2022.2065216}{Adv. Phys.: X \textbf{7,} 2065216 (2022)}.}
	
\bibitem{Sitte2012PRL}{M. Sitte, A. Rosch, E. Altman, and L. Fritz, Topological insulators in magnetic fields: quantum Hall effect and edge channels with a nonquantized $\theta$ term, \href{https://doi.org/10.1103/PhysRevLett.108.126807}{Phys. Rev. Lett. \textbf{108,} 126807 (2012)}.}
	
\bibitem{Zhang2013PRL}{F. Zhang, C. L. Kane, and E. J. Mele, Surface state magnetization and chiral edge states on topological insulators, \href{https://doi.org/10.1103/PhysRevLett.110.046404}{Phys. Rev. Lett. \textbf{110,} 046404 (2013)}.}
	
\bibitem{Benalcazar2017Science}{ W. A. Benalcazar, B. A. Bernevig, and T. L. Hughes, Quantized electric multipole insulators, \href{https://doi.org/10.1126/science.aah644}{Science \textbf{357,} 61 (2017)}.}
	
\bibitem{Schindler2018SciAdv}{F. Schindler, A. M. Cook, M. G. Verginiory, Z. Wang, S. S. P. Parkin, B. A. Bernevig, Higher-order topological insulators, \href{https://doi.org/10.1126/sciadv.aat0346}{Sci. Adv. \textbf{4,} eaat0346(2018)}.}
	
\bibitem{Yang2024JPCM}{Y.-B. Yang, J.-H. Wang, K. Li and Y. Xu, Higher-order topological phases in crystalline and non-crystalline systems: a review, \href{https://doi.org/10.1088/1361-648X/ad3abd}{J. Phys.: Condens. Matter \textbf{36,} 283002 (2024)}.}

\bibitem{Gong2018PRX}{Z. Gong, Y. Ashida, K. Kawabata, K. Takasan, S. Higashikawa, and M. Ueda, Topological phases of non-Hermitian systems, \href{https://doi.org/10.1103/PhysRevX.8.031079}{Phys. Rev. X \textbf{8,} 031079 (2018).}}
	
\bibitem{Bergholtz2021RMP}{E. J. Bergholtz, J. C. Budich, and F. K. Kunst, Exceptional topology of non-Hermitian systems, \href{https://doi.org/10.1103/RevModPhys.93.015005}{Rev. Mod. Phys. \textbf{93,} 015005 (2021)}.}
	
\bibitem{Banerjee2023JPCM}{A. Banerjee, R. Sarkar, S. Dey and A. Narayan, Non-Hermitian topological phases: principles and prospects, \href{https://doi.org/10.1088/1361-648X/acd1cb}{J. Phys.: Condens. Matter \textbf{35,} 333001 (2023)}.}
	
\bibitem{Xue2026PRL}{P. Xue, Essay: Topological phases and exceptional points in non-Hermitian systems, \href{https://doi.org/10.1103/ll76-j2l5}{Phys. Rev. Lett. \textbf{136,} 170001 (2026)}.}
	
\bibitem{Anderson1958}{P. W. Anderson, Absence of diffusion in certain random lattices, \href{https://doi.org/10.1103/PhysRev.109.1492}{Phys. Rev. \textbf{109,} 1492 (1958)}.}
	
\bibitem{Li2009PRL}{J. Li, R.-L. Chu, J. K. Jain, and S.-Q. Shen, Topological Anderson insulator, \href{https://doi.org/10.1103/PhysRevLett.102.136806}{Phys. Rev. Lett. \textbf{102,} 136806 (2009).}}
	
\bibitem{Groth2009PRL}{C. W. Groth, M. Wimmer, A. R. Akhmerov, J. Tworzydło, and C. W. J. Beenakker, Theory of the topological Anderson insulator, \href{https://doi.org/10.1103/PhysRevLett.103.196805}{Phys. Rev. Lett. \textbf{103,} 196805 (2009).}}
	
\bibitem{Xing2011PRB}{Y. Xing, L. Zhang, and J. Wang, Topological Anderson insulator phenomena, \href{https://doi.org/10.1103/PhysRevB.84.035110}{Phys. Rev. B \textbf{84,} 035110 (2011).}}
	
\bibitem{Jiang2009PRB}{H. Jiang, L. Wang, Q.-F. Sun, and X. C. Xie, Numerical study of the topological Anderson insulator in HgTe/CdTe quantum wells, \href{https://doi.org/10.1103/PhysRevB.80.165316}{Phys. Rev. B \textbf{80,} 165316 (2009).}}
	
\bibitem{Zhang2012PRB}{Y.-Y. Zhang, R.-L. Chu, F.-C. Zhang, and S.-Q. Shen, Localization and mobility gap in the topological Anderson insulator, \href{https://doi.org/10.1103/PhysRevB.85.035107}{Phys. Rev. B \textbf{85,} 035107 (2012).}}
	
\bibitem{Song2012PRB}{J. Song, H. Liu, H. Jiang, Q.-F. Sun, and X. C. Xie, Dependence of topological Anderson insulator on the type of disorder, \href{https://doi.org/10.1103/PhysRevB.85.195125}{Phys. Rev. B \textbf{85,} 195125 (2012)}.}
	
\bibitem{Atland2014PRL}{A. Atland, D. Bagrets, L. Fritz, A. Kamenev, and H. Schmiedt, Quantum criticality of quasi-one-dimensional topological Anderson insulators, \href{https://doi.org/10.1103/PhysRevLett.112.206602}{Phys. Rev. Lett. \textbf{112,} 206602 (2014).}}

\bibitem{Shem2014PRL}{I. Mondragon-Shem, T. L. Hughes, J. Song, and E. Prodan, Topological criticality in the chiral-symmetric AIII class at strong disorder, \href{https://doi.org/10.1103/PhysRevLett.113.046802}{Phys. Rev. Lett. \textbf{113,} 046802 (2014).}}
	
\bibitem{Zhang2019PRB}{Z.-Q. Zhang, B.-L. Wu, J. Song, and H. Jiang, Topological Anderson insulator in electric circuits, \href{https://doi.org/10.1103/PhysRevB.100.184202}{Phys. Rev. B \textbf{100,} 184202 (2019).}}
	
\bibitem{Hsu2020PRB}{H.-C. Hsu and T.-W. Chen, Topological Anderson insulating phases in the long-range Su-Schrieffer-Heeger model, \href{https://doi.org/10.1103/PhysRevB.102.205425}{Phys. Rev. B \textbf{102,} 205425 (2020).}}
	
\bibitem{Velury2021PRB}{S. Velury, B. Bradlyn, and T. L. Hughes, Topological crystalline phases in a disordered inversion-symmetric chain, \href{https://doi.org/10.1103/PhysRevB.103.024205}{Phys. Rev. B \textbf{103,} 024205 (2021).}}
	
\bibitem{Zhang2021PRB}{G.-Q. Zhang, L.-Z. Tang, L.-F. Zhang, D.-W. Zhang, and S.-L. Zhu, Connecting topological Anderson and Mott insulators in disordered interacting fermionic systems, \href{https://doi.org/10.1103/PhysRevB.104.L161118}{Phys. Rev. B \textbf{104,} L161118 (2021).}}
	
\bibitem{Lu2022AnndePhys}{Z. Lu, Z. Xu, and Y. Zhang, Exact mobility edges and topological Anderson insulating phase in a slowly varying quasiperiodic model, \href{ https://doi.org/10.1002/andp.202200203}{Ann. Phys. (Berlin) \textbf{534,} 2200203, (2022).}}
	
\bibitem{Ji2025arxiv}{R. Ji, Y. Zhang, S. Chen, and Z. Xu, Multiple re-entrant topological windows induced by generalized Bernoulli disorder, \href{https://doi.org/10.48550/arXiv.2512.06851}{arXiv:2512.06851.}}

\bibitem{Mannai2026arxiv}{M. Manna\"i, Y. Shu, S. Haddad, M. Ren, H. Chen, Y. Sun, and H. Sati, Realization of staircase topological Anderson phase transitions, \href{https://doi.org/10.48550/arXiv.2601.14769}{arXiv:2601.14769}.}
	
\bibitem{Liu2018PLA}{T. Liu and H. Guo, Topological phase transition in the quasiperiodic disordered Su-Schrieffer-Heeger chain, \href{https://doi.org/10.1016/j.physleta.2018.09.023}{Physics Letters A \textbf{382,} 3287 (2018).}}
	
\bibitem{Longhi2020OL}{S. Longhi, Topological Anderson phase in quasi-periodic waveguide lattices, \href{https://doi.org/10.1364/OL.399742}{Optics Letters \textbf{45,} 4036 (2020).}}
	
\bibitem{Tang2022PRA}{L.-Z. Tang, S.-N. Liu, G.-Q. Zhang, and D.-W. Zhang, Topological Anderson insulators with different bulk states in quasiperiodic chains, \href{https://doi.org/10.1103/PhysRevA.105.063327}{Phys. Rev. A \textbf{105,} 063327 (2022).}}
	
\bibitem{Li2024PRR}{X. Li, H. Xu, J. Wang, L.-Z. Tang, D.-W. Zhang, C. Yang, T. Su, C. Wang, Z. Mi, W. Sun, et al., Mapping the topology-localization phase diagram with quasiperiodic disorder using a programmable superconducting simulator, \href{https://doi.org/10.1103/PhysRevResearch.6.L042038}{Phys. Rev. Research \textbf{6,}  L042038 (2024).}}
	
\bibitem{Sircar2025PLA}{S. Sircar, Topological Anderson insulator phases in one dimensional quasi-periodic mechanical SSH chains, \href{https://doi.org/10.1016/j.physleta.2025.130314}{Physics Letters A \textbf{537,} 130314 (2025).}}
	
\bibitem{Wang2026arxiv}{X. Wang, L. Wang, and S. Chen, Topological Anderson insulator and reentrant topological transitions in a mosaic trimer lattice, \href{https://doi.org/10.48550/arXiv.2601.13760}{arXiv:2601.13760.}}

\bibitem{Meier2018Science}{E. J. Meier, F. A. An, A. Dauphin, M. Maffei, P. Massignan, T. L. Hughes, and B. Gadway, Observation of the topological Anderson insulator in disordered atomic wires, \href{https://doi.org/10.1126/science.aat3406}{Science \textbf{362,} 929 (2018)}.}

\bibitem{Stutzer2018Nature}{S. St\"utzer, Y. Plotnik, Y. Lumer, P. Titum, N. H. Lindner, M. Segev, M. C. Rechtsman, and A. Szameit, Photonic topological Anderson insulators, \href{https://doi.org/10.1038/s41586-018-0418-2}{Nature \textbf{560,} 461 (2018)}.}

\bibitem{Liu2020PRL}{G.-G. Liu, Y. Yang, X. Ren, H. Xue, X. Lin, Y.-H. Hu, H.-x. Sun, B. Peng, P. Zhou, Y. Chong, and B. Zhang, Topological Anderson insulator in disordered photonic crystals, \href{https://doi.org/10.1103/PhysRevLett.125.133603}{Phys. Rev. Lett. \textbf{125,} 133603 (2020).}}

\bibitem{Cui2022PRL}{X. Cui, R.-Y. Zhang, Z.-Q. Zhang, and C. T. Chan, Photonic $Z_2$ topological Anderson insulators, \href{https://doi.org/10.1103/PhysRevLett.129.043902}{Phys. Rev. Lett. \textbf{129,} 043902 (2022)}.}

\bibitem{Chen2024PRL}{X.-D. Chen, Z.-X. Gao, X. Cui, H.-C. Mo, W.-J. Chen, R.-Y. Zhang, C. T. Chan, and J.-W. Dong, Realization of time-reversal invariant photonic topological Anderson insulators, \href{https://doi.org/10.1103/PhysRevLett.133.133802}{Phys. Rev. Lett. \textbf{133,} 133802 (2024).}}

\bibitem{Ren2024PRL}{M. Ren, Y. Yu, B. Wu, X. Qi, Y. Wang, X. Yao, J. Ren, Z. Guo, H. Jiang, H. Chen, X.-J. Liu, Z. Chen, and Y. Sun, Realization of gapped and ungapped photonic topological Anderson insulators, \href{https://doi.org/10.1103/PhysRevLett.132.066602}{Phys. Rev. Lett. \textbf{132,} 066602 (2024).}}

\bibitem{Nejad2022AM}{F. Zangeneh-Nejad and R. Fleury, Disorder-induced signal filtering with topological metamaterials, \href{https://doi.org/10.1002/adma.202001034}{Adv. Mater. \textbf{32,} 2001034 (2020)}.}

\bibitem{Gu2023SCPMA}{Z. Gu, H. Gao, H. Xue, D. Wang, J. Guo, Z. Su, B. Zhang, and J. Zhu, Observation of an acoustic non-Hermitian topological Anderson insulator, \href{https://doi.org/10.1007/s11433-023-2159-4}{Sci. China Phys. Mech. Astron. \textbf{66,} 294311 (2023)}.}

\bibitem{Zhang2021PRL}{W. Zhang, D. Zou, Q. Pei, W. He, J. Bao, H. Sun, and X. Zhang, Experimental observation of higher-order topological Anderson insulators, \href{https://doi.org/10.1103/PhysRevLett.126.146802}{Phys. Rev. Lett. 126, 146802 (2021)}.}
	
\bibitem{Zhang2020SciChina}{D.-W. Zhang, L.-Z. Tang, L.-J. Lang, H. Yan, and S.-L. Zhang, Non-Hermitian topological Anderson insulators, \href{https://doi.org/10.1007/s11433-020-1521-9}{Sci. China-Phys. Mech. Astron. \textbf{63,} 267062 (2020).}}
	
\bibitem{Liu2020CPB}{H. Liu, Z. Su, Z.-Q. Zhang, and H. Jiang, Topological Anderson insulator in two-dimensional non-Hermitian systems, \href{https://doi.org/10.1088/1674-1056/ab8201}{Chinese Phys. B \textbf{29,} 050502 (2020).}}
	
\bibitem{Tang2020PRA}{L.-Z. Tang, L.-F. Zhang, G.-Q. Zhang, and D.-W. Zhang, Topological Anderson insulators in two-dimensional non-Hermitian disordered systems, \href{https://doi.org/10.1103/PhysRevA.101.063612}{Phys. Rev. A \textbf{101,} 063612 (2020)}}
	
\bibitem{Lin2022NatCom}{Q. Lin, T. Li, L. Xiao, K. Wang, W. Yi, and P. Xue, Observation of non-Hermitian topological Anderson insulator in quantum dynamics, \href{https://doi.org/10.1038/s41467-022-30938-9}{Nat. Commun. \textbf{13,} 3229 (2022).}}

\bibitem{Zeng2026PRB}{Q.-B. Zeng and R. L\"u, Coexistence of topological Anderson insulator and multifractal critical phase in a non-Hermitian quasicrystal, \href{https://doi.org/10.1103/n9rf-zk9t}{Phys. Rev. B \textbf{113,} 224203 (2026)}.}

\bibitem{Tezuka2012PRB}{M. Tezuka and N. Kawakami, Reentrant topological transitions in a quantum wire/superconductor system with quasiperiodic lattice modulation, \href{https://doi.org/10.1103/PhysRevB.85.140508}{Phys. Rev. B \textbf{85,} 140508(R) (2012)}.}

\bibitem{Beugeling2012PRB}{W. Beugeling, C. X. Liu, E. G. Novik, L. W. Molenkamp, and C. Morais Smith, Reentrant topological phases in Mn-doped HgTe quantum wells, \href{https://doi.org/10.1103/PhysRevB.85.195304}{Phys. Rev. B \textbf{85,} 195304 (2012)}.}

\bibitem{Rieder2013PRB}{M.-T. Rieder, P. W. Brouwer, and \.I. Adagideli, Reentrant topological phase transitions in a disordered spinless superconducting wire, \href{https://doi.org/10.1103/PhysRevB.88.060509}{Phys. Rev. B \textbf{88,} 060509 (2013)}.}

\bibitem{Tezuka2013PRB}{M. Tezuka and N. Kawakami, Reentrant topological transitions with majorana end states in one-dimensional superconductors by lattice modulation, \href{https://doi.org/10.1103/PhysRevB.88.155428}{Phys. Rev. B \textbf{88,} 155428 (2013)}.}

\bibitem{Hsu2017PRB}{H.-C. Hsu, M.-J. Jhang, T.-W. Chen, and G.-Y. Guo, Topological phase transitions in an inverted InAs/GaSb quantum well driven by tilted magnetic fields, \href{https://doi.org/10.1103/PhysRevB.95.195408}{Phys. Rev. B \textbf{95,} 195408 (2017)}.}

\bibitem{Sugimoto2017PRB}{T. Sugimoto, M. Ohtsu, and T. Tohyama, Reentrant topological phase transition in a bridging model between Kitaev and Haldane chains, \href{https://doi.org/10.1103/PhysRevB.96.245118}{Phys. Rev. B \textbf{96,} 245118 (2017)}.}

\bibitem{Yang2020PRB}{Y. Yang, S.-J. Ran, X. Chen, Z.-Z. Sun, S.-S. Gong, Z. Wang, and G. Su, Reentrance of the topological phase in a spin-1 frustrated heisenberg chain, \href{https://doi.org/10.1103/PhysRevB.101.045133}{Phys. Rev. B \textbf{101,} 045133 (2020)}.}

\bibitem{Padhan2024PRB}{A. Padhan, S. R. Padhi, and T. Mishra, Complete delocalization and reentrant topological transition in a non-Hermitian quasiperiodic lattice, \href{https://doi.org/10.1103/PhysRevB.109.L020203}{Phys. Rev. B \textbf{109,} L020203 (2024)}.}

\bibitem{Kesharpu2025PRB}{K. K. Kesharpu, E. A. Kochetov, and A. Ferraz, Reentrant topological order in a strongly correlated nanowire due to Rashba spin-orbit coupling, \href{https://doi.org/10.1103/PhysRevB.111.115153}{Phys. Rev. B \textbf{111,} 115153 (2025)}.}

\bibitem{Cinnirella2025PRB}{E. G. Cinnirella, A. Nava, G. Campagnano, and D. Giuliano, Phase diagram of the disordered Kitaev chain with long-range pairing connected to external baths, \href{https://doi.org/10.1103/PhysRevB.111.155149}{Phys. Rev. B \textbf{111,} 155149 (2025)}.}

\bibitem{Li2025PRB}{C.-A. Li, B. Fu, J. Li, and B. Trauzettel, Random-fluxinduced transition sequence between weak and strong topological phases with anisotropic localization properties, \href{https://doi.org/10.1103/zjyw-ln2n}{Phys. Rev. B \textbf{111,} 214207 (2025)}.}

\bibitem{Lu20252025FoP}{Z. Lu, Y. Zhang, and Z. Xu, Reentrant localization transitions in a topological Anderson insulator: A study of a generalized Su-Schrieffer-Heeger quasicrystal, \href{https://doi.org/10.15302/frontphys.2025.024204}{Frontiers of Physics \textbf{20,} 024204 (2025)}.}

\bibitem{Wang2025PRA}{X.-M. Wang, S.-Z. Li, and Z. Li, Emergent topological re-entrant phase transition in a generalized quasiperiodic modulated Su-Schrieffer-Heeger model, \href{https://doi.org/10.1103/PhysRevA.111.022214}{Phys. Rev. A \textbf{111,} 022214 (2025)}.}

\bibitem{Sinha2025PRA}{A. Sinha, T. Shit, A. Tetarwal, D. Sen, and S. Mukherjee, Probing the topological anderson transition in quasiperiodic photonic lattices via chiral displacement and wavelength tuning, \href{https://doi.org/10.1103/9jjd-vbp1}{Phys. Rev. A \textbf{112,} 013512 (2025)}.}

\bibitem{Song2014PRB}{J. Song and E. Prodan, AIII and BDI topological systems at strong disorder, \href{https://doi.org/10.1103/PhysRevB.89.224203}{Phys. Rev. B \textbf{89,} 224203 (2014)}.}
	
\end{thebibliography}
\end{document}